\documentclass[conference]{IEEEtran}
\IEEEoverridecommandlockouts
\usepackage{cite}
\usepackage{amsmath,amssymb,amsfonts}
\usepackage{graphicx}
\usepackage{textcomp}
\usepackage{xcolor}
\usepackage[utf8]{inputenc}
\usepackage[T1]{fontenc}
\usepackage{times}
\usepackage{booktabs}
\usepackage{tabularx}
\usepackage{enumitem}
\usepackage{hyperref}

\hypersetup{
  colorlinks=true,
  linkcolor=blue!60!black,
  citecolor=blue!60!black,
  urlcolor=blue!60!black
}

\def\BibTeX{{\rm B\kern-.05em{\sc i\kern-.025em b}\kern-.08em
    T\kern-.1667em\lower.7ex\hbox{E}\kern-.125emX}}

\begin{document}
\bstctlcite{IEEEexample:BSTcontrol} 

\title{SubEdge: A Subscriber-Centric Edge Computing Subsystem in 6G Networks for AI}

\author{
\IEEEauthorblockN{Abdirazak Ali Asir Rage, Riccardo Pozza, and Rahim Tafazolli}
\IEEEauthorblockA{
6G Innovation Centre (6GIC), Institute for Communication Systems, University of Surrey, Guildford, United Kingdom \\
Email: \{abdirazak.rage, r.pozza, r.tafazolli\}@surrey.ac.uk}
}

\maketitle

\begin{abstract}

Beyond traditional connectivity, 6G is envisioned to transform mobile networks into a distributed fabric that provides native integrated communication, computing, and intelligence services. AI-native terminals (e.g., robots, autonomous vehicles, and smart glasses) require real-time inference from individualised, manufacturer-specific models that cannot be executed on-board nor shared across subscribers, making per-subscriber edge compute the necessary complement to per-subscriber connectivity. Existing Network for AI (Net4AI) architectures provision compute for application providers through shared deployments and do not address per-subscriber provisioning. This paper proposes \textbf{SubEdge}, a Net4AI subsystem that provisions integrated communication and compute resources on a per-subscriber basis, ensuring the coupled migration of both dimensions to maintain service continuity during mobility. SubEdge contributes the \emph{computing context}---a per-subscriber data structure binding a Subscription Permanent Identifier (SUPI) to its inference container, edge node, and service entitlement---and a mobility-event-driven mechanism that simultaneously
migrates the subscriber's compute instance and its traffic-routing policy when the serving cell changes. SubEdge operates as an Application Function over existing Network Exposure Function (NEF) APIs with zero 3GPP core modifications. Experimental evaluation on a real-world testbed shows that SubEdge's mobility-driven joint communication-and-compute migration reduces 95th-percentile latency from 22.9\,ms to 12.2\,ms with zero packet loss across six mobility events, sustains 99.92\% frame delivery for an end-to-end 30\,fps inference workload, and completes 1{,}560 migration operations across batches of up to 50 simultaneously migrating subscribers with 100\% success.
\end{abstract}


\section{Introduction}

AI-native terminals (e.g., robots, autonomous vehicles, and smart glasses) increasingly depend on real-time inference from manufacturer-specific, individualised AI models for navigation, perception, or scene
understanding. Such inference cannot be executed
on-board: the graphics processing unit (GPU)
hardware required renders the
terminal prohibitively expensive, heavy, and
power-hungry, with a compute gap of an order of
magnitude constrained by physics rather than
technology~\cite{reuther2020survey}. AI-native terminals must
therefore offload inference to the network, where
dedicated compute is provisioned \emph{per
subscriber}: each model is fine-tuned to that
device's sensors, operational context, and task
profile~\cite{shuvo2022efficient}, and cannot be shared the way a
content delivery network (CDN) cache or
cloud-gaming server can. Cloud inference offers
sufficient capacity but its 40--200\,ms round-trip
latency exceeds the 33\,ms frame budget for real-time
perception at 30\,fps, ruling it out for
latency-critical workloads such as robot navigation.
Mobile operators, uniquely positioned with edge nodes
at cell sites delivering 5--20\,ms latency, are the
natural providers of this per-subscriber inference
compute. This requirement is consistent with the
broader 6G architectural commitment to integrated
communication and computing, in which the
next-generation mobile network is framed not merely
as a connectivity substrate but as a native provider
of compute services with jointly managed
end-to-end quality of service~\cite{yuan20256g,peng20256g,tenny2025toward}.

The Net4AI agenda has produced a body of
architectures targeting this challenge: Multi-Access Edge Computing (MEC)~\cite{etsi_mec003}, the 3GPP
edge-application framework
(EDGEAPP)~\cite{tgpp_23558}, In-Network
Computing~\cite{guerzoni2026network}, network-slicing
proposals~\cite{moreira2025unleashing,wu2022ai,nadar2024enhancing}, and AI
Sessions~\cite{chraiti2026ai}. Each contributes a
piece of the puzzle; yet all share a fundamental
design mismatch with the per-subscriber requirement,
in that they provision compute for \emph{application
providers} deploying \emph{shared} services rather
than for individual subscribers who each register
their own dedicated inference workload. None binds
edge resources to a Subscription Permanent
Identifier (SUPI), none specifies how a
subscriber-owned workload enters operator
infrastructure, and none treats 3GPP subscriber
mobility events as compute lifecycle triggers. These
architectures were designed for human smartphone
users consuming shared applications; the AI-terminal
subscriber, who \emph{owns} a personalised model and
requires dedicated edge execution, represents a
fundamentally new subscriber class that no existing
proposal addresses.

This paper proposes \textbf{SubEdge}, a subsystem
that enables mobile operators to provision
individualised inference compute to their AI-terminal
subscribers alongside connectivity. SubEdge treats
each subscriber's inference container as a
first-class network object bound to their SUPI. Its
central design principle is that this
provisioning is \emph{inseparable} from mobility
management: as the AI terminal moves, the operator
must simultaneously relocate the inference container
to the nearer edge node (\emph{compute migration})
and reroute inference traffic to follow it
(\emph{communication migration}). No existing Net4AI
standard binds compute to subscriber identity at
this granularity, and consequently none drives
migration directly from 3GPP subscriber mobility
events. SubEdge deploys as an Application Function
(AF) over existing Network Exposure Function (NEF)
application programming interfaces (APIs), following
the IP Multimedia Subsystem (IMS) pattern: a
subsystem that introduces subscriber-bound service
management without modifying the underlying 3GPP
core. The contributions of this work are as follows.

\begin{enumerate}[leftmargin=*,nosep]
\item We define the \emph{computing context}: a
per-subscriber data structure binding SUPI to a
container instance, edge node, workload reference,
execution state, and service tier. It is to
subscriber compute what the Protocol Data Unit (PDU)
session is to subscriber connectivity.

\item We design a \emph{subscriber mobility-driven
migration mechanism} that coordinates compute and
communication simultaneously: SubEdge redeploys the
inference container on the nearer edge node and
updates the \texttt{Nnef\_TrafficInfluence} routing
policy in a single NEF call, transparent to the
terminal and the model provider.

\item We implement and evaluate SubEdge on unmodified
free5GC v4.0.1 across three complementary dimensions:
link-layer 95th-percentile round-trip time (p95 RTT)
reduced from 22.9\,ms to 12.2\,ms with zero packet
loss; application-layer 99.92\% frame delivery for
an end-to-end 30\,fps inference workload across the
same mobility scenario; and control-plane 100\%
success across 1{,}560 concurrent migration
operations for $N \in \{1, 5, 10, 20, 30, 40, 50\}$
subscribers.
\end{enumerate}

The remainder of this paper is organised as follows.
Section~\ref{sec:related} reviews existing Net4AI
architectures and identifies the gaps that motivate a
subscriber-centric design. Section~\ref{sec:arch}
presents the SubEdge architecture, detailing its six
internal components, the computing context primitive, and
the coordinated compute-and-communication migration
mechanism. Section~\ref{sec:performance_eval} reports
the experimental evaluation on a real-world testbed
running unmodified free5GC v4.0.1 across link,
application, and control-plane dimensions, and
discusses the scope limitations of the current
testbed. Section~\ref{sec:conclusion} concludes and
outlines directions for future work.

\section{Related Work}
\label{sec:related}

The literature on network support for AI workloads spans
three categories: edge platforms, in-network compute, and
session-layer abstractions.

ETSI MEC~\cite{etsi_mec003} and 3GPP EDGEAPP~\cite{tgpp_23558} are
designed around the application provider as the
central actor and target shared-application scenarios
in which a single instance serves many subscribers.
Authors in~\cite{peng20256g} show empirically that the
overlay fusion model underlying both standards keeps
communication and compute under separate control,
incurring 88\,ms task delay and over 573\,ms task
jitter under realistic mobility. Neither standard
defines a per-subscriber data structure binding a
SUPI to a container instance: the subscriber is a
traffic source, not the principal of the compute
lifecycle.

A subsequent generation of proposals natively fuses
communication and computing inside the operator
core. Authors in~\cite{peng20256g} introduce the Mobile
Computing-Aware Network (MCAN) with dual-functional
Computing Executors, reporting 62\% delay and 99\%
jitter reductions versus 5G MEC. Authors
in~\cite{tenny2025toward} couple compute and transport via a
joint latency budget and improve service capacity by
60\%. Authors in~\cite{yuan20256g} integrate a Computing
Control Function with internal AI nodes and
demonstrate 64--95\% reduction in uplink processing
delay. Authors in~\cite{guerzoni2026network}
and~\cite{schwarzmann2024native} respectively
propose an architectural framework for in-network
computing in the 6G core and an AI-execution-capable
User Plane Function (UPF). Together, these proposals
establish the quantitative case for native
communication-compute integration. Their unit of
provisioning, however, is the application service
graph rather than the subscriber: compute placement
is driven by application-level optimisation, the AI
workload is treated as a shared service, and most
require modification to standardised 3GPP network
functions.

The closest prior work to SubEdge is the AI Sessions
(AIS) in~\cite{chraiti2026ai},
which defines a contractual lifecycle object binding
model identity, execution placement, transport
quality of service, and consent/charging scope,
jointly co-reserved via PREPARE/COMMIT and
re-bound via make-before-break. AIS re-anchors a
session when its analytics predictor signals risk of
service-level violation, regardless of user mobility, and does not specify a
per-subscriber dedicated compute instance hosted on
an edge node selected for proximity to the
subscriber's serving cell that follows the
subscriber under mobility. SubEdge provisions
exactly that: integrated communication and compute
per subscriber, jointly migrated under 3GPP
subscriber mobility events. As the UE moves, SubEdge
selects a nearby edge node, migrates the AI model
container to that node, and routes the subscriber's
inference traffic to it. Authors
in~\cite{moreira2025unleashing,wu2022ai,nadar2024enhancing} treat AI as a traffic
category managed via slicing and the Network Data
Analytics Function (NWDAF); these proposals bind
compute to slice or application identifiers rather
than to subscribers.

\begin{table}[t]
\centering
\caption{Capability comparison of Net4AI proposals.
``\checkmark'' specified; ``$\circ$'' partial;
``--'' absent.}
\label{tab:gap}
\renewcommand{\arraystretch}{1.15}
\setlength{\tabcolsep}{2.5pt}
\scriptsize
\begin{tabular}{@{}p{0.4\columnwidth}ccccc@{}}
\toprule
Capability & MEC & EDGE- & MCAN/ & AIS & SubEdge \\
           & \cite{etsi_mec003} & APP\,\cite{tgpp_23558} & INC\,\cite{peng20256g,guerzoni2026network} & \cite{chraiti2026ai} & (This paper)\\
\midrule
Edge-proximity execution         & \checkmark & \checkmark & \checkmark & \checkmark & \checkmark \\
Compute--transport co-reserve.    & --        & --        & \checkmark & \checkmark & \checkmark \\
Per-subscriber compute provisioning     & --        & --        & --        & --        & \checkmark \\
Joint compute and communication migration.     & --        & --        & --        & $\circ$   & \checkmark \\
Mobility event-driven trigger    & $\circ$   & $\circ$   & --        & --        & \checkmark \\
\bottomrule
\end{tabular}
\end{table}

A comparison of related work is presented in Table \ref{tab:gap}. In summary, no prior study has proposed integrated communication and compute per subscriber with joint communication-and-compute migration triggered by
3GPP subscriber mobility events. SubEdge is designed
to fill this gap as an Application Function over
existing NEF APIs, requiring no modification to
standardised 3GPP network functions.

\section{SubEdge Architecture}
\label{sec:arch}

\begin{figure}[t]
  \centering
  \includegraphics[width=\columnwidth]{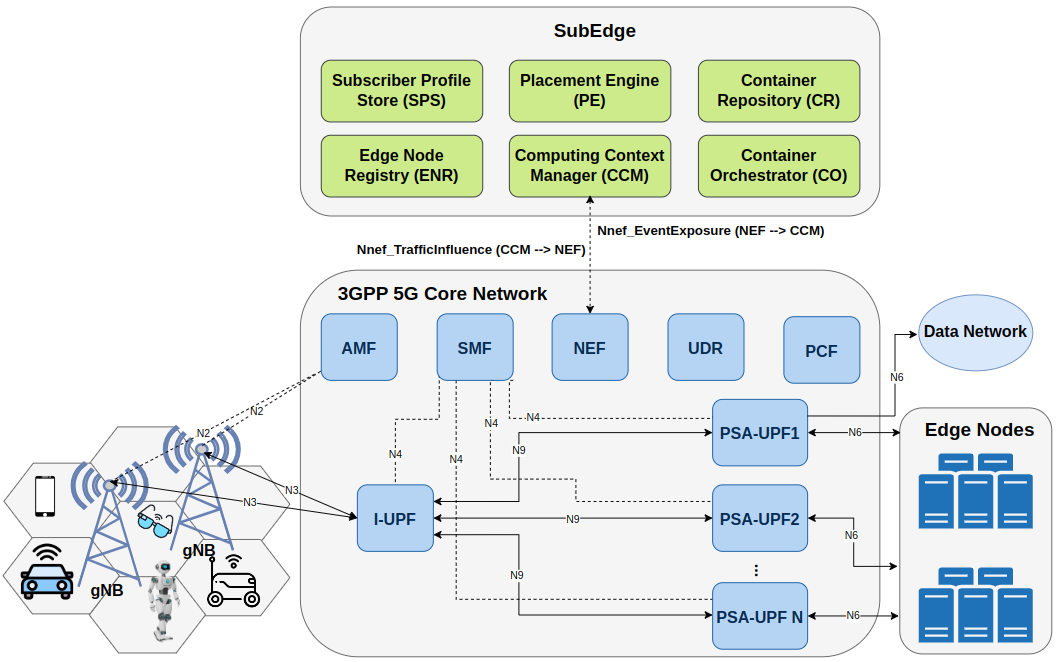}
  \caption{SubEdge architecture and 3GPP integration.
    Six internal components (CCM, SPS, PE, CR, CO, ENR)
    interact with the 5GC exclusively via NEF using
    \texttt{Nnef\_EventExposure} and
    \texttt{Nnef\_TrafficInfluence}. Mobility events
    trigger the standard policy chain
    (NEF\,$\to$\,UDR\,$\to$\,PCF\,$\to$\,SMF\,$\to$\,UPF)
    that redirects inference traffic. No core network
    function requires modification.}
  \label{fig:arch}
\end{figure}

SubEdge interfaces with the 3GPP core exclusively
through the NEF, consuming two service operations
defined in TS\,23.502~\cite{3gppTS23.502} and TS\,29.522~\cite{3gppTS29.522}.
\texttt{Nnef\_EventExposure} carries subscriber
lifecycle events from the Access and Mobility
Management Function (AMF) through the NEF to the
SubEdge Computing Context Manager (CCM): PDU
session establishment triggers context creation and
container deployment; handover notification triggers
placement re-evaluation and potential migration;
session release triggers context suspension. \texttt{Nnef\_TrafficInfluence} allows
SubEdge to request per-subscriber routing policy by
supplying a Data Network Access Identifier (DNAI) pointing to
the target edge node. The resulting policy
chain---the NEF persists the AF influence data in the
Unified Data Repository (UDR); the Policy Control
Function (PCF) derives Policy and Charging Control
(PCC) rules; the Session Management Function (SMF)
enforces them on the UPF via the N4 reference
point---is the standard 3GPP traffic steering
mechanism, requiring no modification to any core
network function.

Six internal components, illustrated in
Fig.~\ref{fig:arch}, implement the subscriber-centric
compute lifecycle.

The \textbf{Computing Context Manager (CCM)} is the
central orchestrator and the sole component that
interfaces with NEF. For each active subscriber the CCM
maintains a \emph{computing context} record containing
the SUPI, context state (created / active / migrating /
suspended / deleted), assigned edge node, container
image reference, service tier, current serving cell ID,
and the active TrafficInfluence subscription reference.
This record is the architectural primitive that
distinguishes SubEdge: no current specification binds a
compute instance to a subscriber identity at this
granularity, and its absence from existing proposals is
the root cause of the gaps identified in
Section~\ref{sec:related}.

The \textbf{Subscriber Profile Store (SPS)} maps each
SUPI to its compute service entitlement---tier
(basic / standard / premium), GPU allocation,
service-level agreement (SLA) on latency, execution
mode (hot / warm / cold), and migration
priority---governing every downstream placement
decision and mapping directly to operator
business-to-consumer (B2C) plans for per-subscriber
GPU billing through the Charging Function (CHF).

The \textbf{Placement Engine (PE)}, drawing on the
SPS and the Edge Node Registry, selects the edge
node that hosts each subscriber's AI model container
given their serving cell and resource requirements.
The PE estimates the path latency between the
subscriber's serving cell and each candidate edge
node, using the Edge Node Registry's cell-to-node
mapping, and selects the node that minimises this
latency subject to the subscriber's GPU and
service-level requirements. It applies selective
re-evaluation during mobility events: if the current
node remains within the subscriber's latency SLA
after a handover, no migration is triggered,
avoiding container churn in dense urban deployments
where a single edge node covers many adjacent cells.

The \textbf{Container Repository (CR)} stores,
validates, and distributes subscriber-owned Open
Container Initiative (OCI) container images. The
primary channel is manufacturer backend integration,
mirroring how device manufacturers certify handsets
for operator networks, providing a workload ingestion path absent from existing edge
architectures. The CR maintains a distributed cache
across edge nodes; pre-staged images eliminate the
remote pull from the migration critical path,
transforming migration from a cold pull taking tens
of seconds into a warm local start taking
seconds, directly addressing the feasibility
concern noted in~\cite{tenny2025toward} regarding the latency
cost of workload redistribution.

The \textbf{Container Orchestrator (CO)} executes the
container lifecycle on edge nodes, translating CCM
instructions into Kubernetes operations: image
pulls, pod creation with GPU and CPU limits derived
from the SPS, health checks, make-before-break
sequencing across source and target nodes during
migration, and teardown after migration completes.
The CO acts on whichever edge node the CCM directs
it to and does not itself decide where workloads run.

The \textbf{Edge Node Registry (ENR)} is the operator-side view of available edge infrastructure and is the sixth and final SubEdge component, distinct from the
CO. Whereas the CO acts on edge nodes, the ENR
\emph{describes} them: it maintains real-time GPU
availability, the cell-ID-to-node mapping, and the
health state of every registered edge node. The PE
consults the ENR to decide where each subscriber's
container should be placed, and the CCM consults it
to detect degraded infrastructure and proactively
migrate affected subscribers. Separating placement
information (ENR) from placement execution (CO) keeps
SubEdge's planning and acting layers cleanly
decoupled.

The SPS specifies one of three execution modes per
subscriber. \emph{Hot} mode (dedicated GPU, always
running) targets robots requiring continuous 30\,fps
inference; \emph{warm} (GPU time-shared) targets smart
glasses; \emph{cold} (started on demand) targets bursty
workloads. Only hot subscribers consume dedicated GPU at
all times; warm and cold tiers increase subscriber
density 3--5$\times$ per node.

\noindent\textbf{Coordinated compute and communication
migration.} The defining behaviour of SubEdge is
the simultaneous coordination of \emph{both}
dimensions of subscriber service when the subscriber
moves---the joint management imperative identified
across~\cite{peng20256g,tenny2025toward,yuan20256g} instantiated at the
per-subscriber level. When the AMF reports a
serving-cell change via
\texttt{Nnef\_EventExposure}, the CCM invokes the PE
to re-evaluate placement for the new serving cell.
If the PE returns a different edge node, the CCM
orchestrates a coordinated two-phase migration. In the
\emph{compute migration} phase, the CCM instructs
the CO to deploy a new container on the nearby edge
node, pulling the pre-staged image from the local
CR cache. In the \emph{communication migration}
phase, executed once the new container is healthy,
the CCM issues a single
\texttt{Nnef\_TrafficInfluence} PUT to the NEF
specifying the target DNAI, triggering the standard
5GC policy chain that redirects the subscriber's
inference traffic to the new container. The source
container is then stopped. The terminal has no
awareness of either phase: it continues sending
inference data to the server's virtual IP address
(VIP) throughout, experiencing no service
interruption because both the execution environment
and the traffic path follow the subscriber by
tracking the nearby edge node selected for the
current serving cell.

\section{Performance Evaluation}
\label{sec:performance_eval}

\subsection{Experimental setup}

\begin{figure}[t]
  \centering
  \includegraphics[width=\columnwidth]{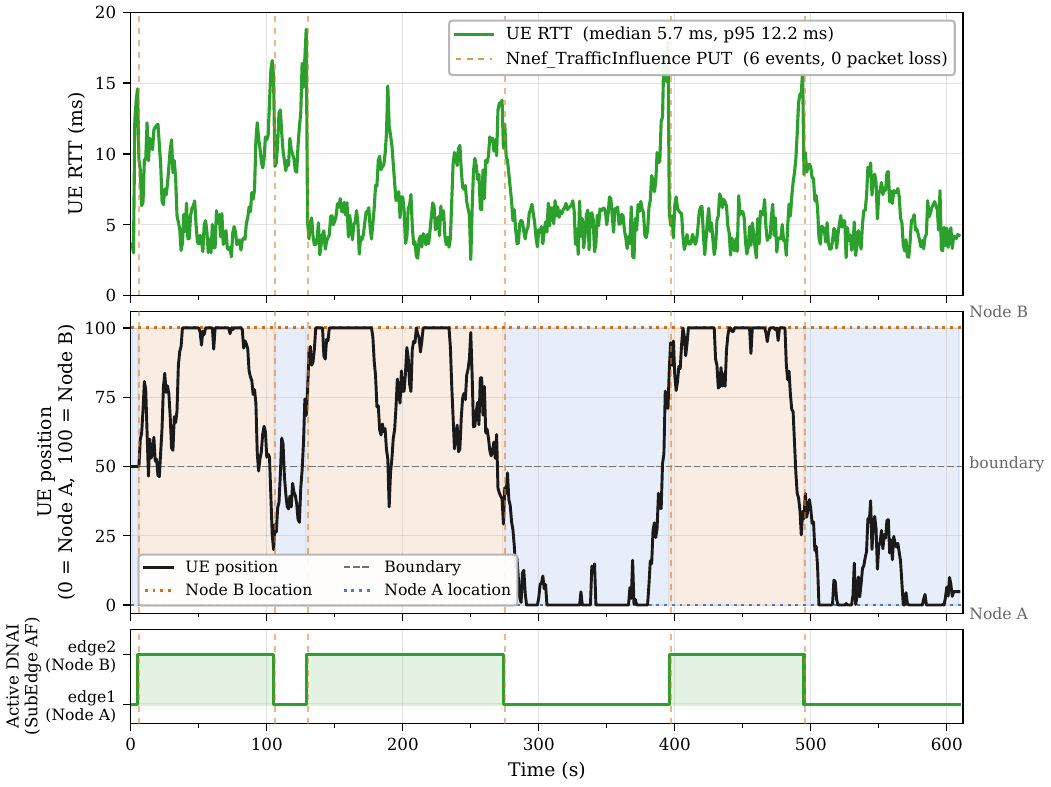}
  \caption{Link-layer view of mobility-driven migration
    over 600\,s. \textit{Top:} UE-observed ICMP RTT
    (1\,Hz); orange dashed lines mark six
    \texttt{Nnef\_TrafficInfluence} PUTs.
    \textit{Middle:} UE position; shading indicates the
    active edge node (blue: A; amber: B). \textit{Bottom:}
    active-DNAI step function. Zero packet loss across
    609 samples.}
  \label{fig:link_layer_latency}
\end{figure}

\begin{figure}[t]
  \centering
  \includegraphics{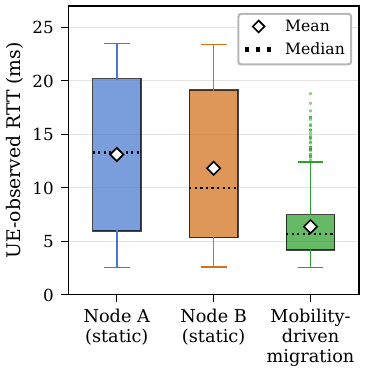}
  \caption{RTT distribution across 607--609 ICMP samples
    (1\,Hz, 0\% loss): static Node~A
    (p95\,=\,22.9\,ms), static Node~B
    (p95\,=\,22.8\,ms), and SubEdge migration
    (p95\,=\,12.2\,ms). Bold labels show the share of
    samples below 10\,ms (41\%, 50\%, 88\%).}
  \label{fig:boxplot_latency}
\end{figure}

The SubEdge AF is implemented in Go across six packages
corresponding to the six architectural components,
consuming free5GC v4.0.1's northbound NEF HTTP/2 API.
The testbed comprises a single host running free5GC
v4.0.1 (AMF, SMF, PCF, UDR, NEF) and two PDU Session
Anchor UPFs (PSA-UPFs), with UERANSIM providing the
simulated UE and two physical edge nodes each
hosting a containerised inference endpoint. Each edge node binds a shared
service VIP on its loopback interface; the active UPF
traffic-steering policy, driven by SubEdge
\texttt{Nnef\_TrafficInfluence} updates, determines
which physical node receives a given UE's traffic. Both
candidate edge nodes run pre-staged inference
containers continuously in hot mode, mirroring the
always-on requirement of robotics and autonomous-driving
subscribers per~\S\,\ref{sec:arch}.

Mobility-driven latency fluctuations provide the migration triggers: a bounded Gaussian random-walk
model (seed 183, std 8.0) drives \texttt{tc netem}
on the GPRS Tunnelling Protocol User-plane (GTP-U)
interfaces of both PSA-UPFs in counter-phase over a
600\,s run, replicating the path quality variation a
mobile subscriber experiences as their proximity to
each edge node changes. The migration algorithm applies a
5\,s hysteresis hold before issuing a
\texttt{Nnef\_TrafficInfluence} PUT, preventing
oscillation at path-quality crossings without RTT cost
since both paths carry comparable latency at crossings
by definition. Because free5GC v4.0.1 does not yet
expose \texttt{Nnef\_EventExposure}, SubEdge monitors
path latency directly and issues PUTs when the active
node's RTT exceeds the SLA threshold. We note that
proximity- and event-triggered migration may differ in
their transient behaviour under abrupt handovers (an
event-driven trigger could fire ahead of measurable
path degradation), but the migration mechanism that the
trigger invokes (the two-phase compute-and-communication
sequence of \S\,\ref{sec:arch}) is identical in both
cases, and substituting the trigger source is a
configuration change once the event becomes available
in free5GC.

The migration mechanism is evaluated through two
independent workloads measuring distinct properties.
The first is a 1\,Hz ICMP probe, isolating the
network-path contribution to round-trip latency
(Figs.~\ref{fig:link_layer_latency}--\ref{fig:boxplot_latency}). The
second is a 30\,Hz HTTP/POST inference workload
(150\,KB request body representing a 224$\times$224
image frame, emulated 5\,ms inference on the
container, $\sim$1\,KB JSON response), running
through the UE's \texttt{uesimtun0} interface and
traversing the full gNB$\to$UPF$\to$edge-node data
path (Fig.~\ref{fig:inference_continuity}). The 5\,ms
emulated inference is a deliberate methodological
choice: real per-frame inference latency depends on the
model architecture, batch size, and GPU generation, and
varies by an order of magnitude across the AI-terminal
classes this work targets. A third experiment validates concurrent migration: $N$
concurrent \texttt{Nnef\_TrafficInfluence} PUT
operations are issued via a \texttt{threading.Barrier},
one per distinct SUPI, for $N \in \{1,5,10,20,30,40,
50\}$, ten repetitions per $N$.

\subsection{Results}

\subsubsection{Link-Layer: Mobility-Driven Latency
Reduction}

\begin{figure}[t]
  \centering
  \includegraphics[width=\columnwidth]{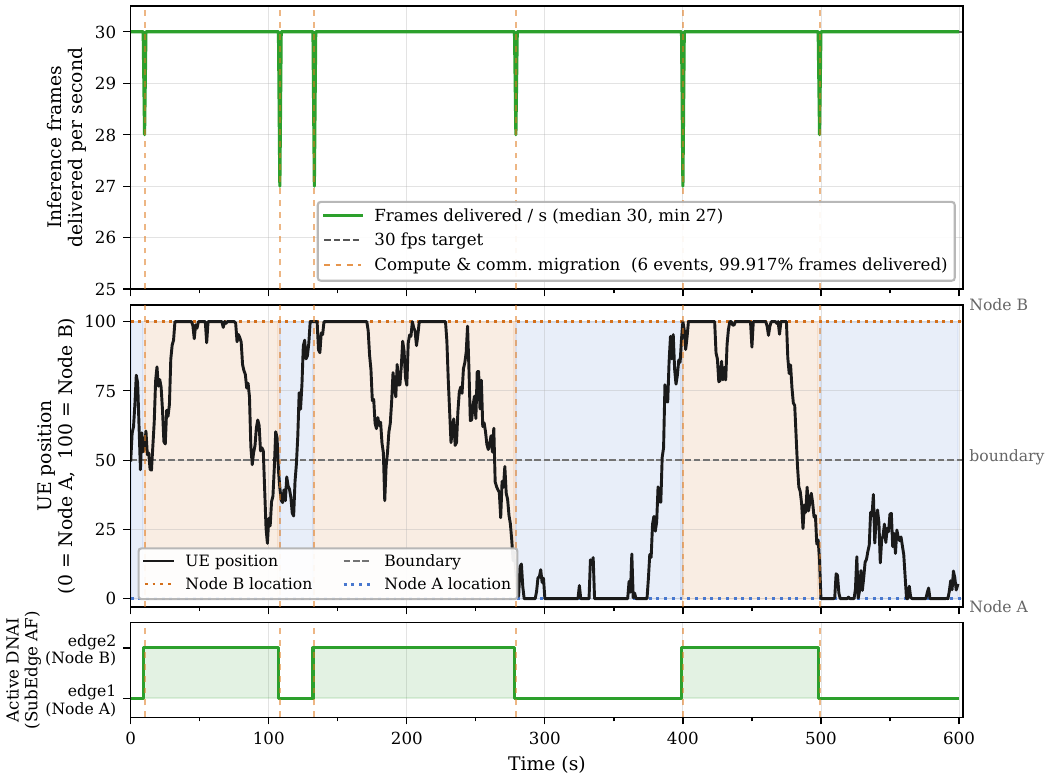}
  \caption{Application-layer view of the same migration
    scenario as Fig.~\ref{fig:link_layer_latency}. \textit{Top:}
    delivered inference frames per 1\,s window (30\,fps
    target, 18{,}000-frame run). Six single-second dips
    to 27--28\,fps coincide with migrations;
    17{,}985/18{,}000 frames delivered (99.92\%).
    \textit{Middle/bottom:} matched UE position and DNAI
    traces.}
  \label{fig:inference_continuity}
\end{figure}

Fig.~\ref{fig:link_layer_latency} presents the headline
link-layer result. The UE traverses the random-walk
trajectory, spending roughly equal time near each edge
node. The background shading---keyed to the active
\texttt{Nnef\_TrafficInfluence} subscription, not to UE
geometric position---flips exactly at each migration
marker, making the causal chain explicit: as the UE
moves closer to the alternative edge node, SubEdge
detects lower path latency for five consecutive seconds,
issues a single PUT to NEF, and the policy chain
(NEF\,$\to$\,UDR\,$\to$\,PCF\,$\to$\,SMF\,$\to$\,UPF~N4)
redirects inference traffic within one ping interval.
The terminal observed none of this: 609 ICMP pings were
transmitted and 609 received, the \texttt{icmp\_seq}
counter contiguous throughout.

Fig.~\ref{fig:boxplot_latency} quantifies the RTT improvement
against static-assignment baselines over matched runs. Static Node~A yields median 13.3\,ms,
p95 22.9\,ms, with only 41\% of samples below 10\,ms.
Static Node~B reduces the median to 10.0\,ms but its
p95 remains at 22.8\,ms because the random walk places
the UE far from Node~B for extended periods; Node~B's
wide IQR reflects this bimodal behaviour. SubEdge
migration achieves \textbf{median 5.7\,ms and p95
12.2\,ms}, with \textbf{88\%} of samples below
10\,ms---a 47\,pp improvement over Node~A and 38\,pp
over Node~B. The SubEdge box sits entirely below both
static IQRs, and its p95 is nearly half that of either
baseline. The gain is not an artefact of spending more
time near one node: the trajectory is balanced (45\%
Node~A zone, 55\% Node~B zone), so the improvement is
attributable entirely to mobility-aware placement.

\subsubsection{Application-Layer: AI Inference Service
Continuity}
\label{sec:inference}

\begin{figure}[t]
  \centering
  \includegraphics[width=\columnwidth]{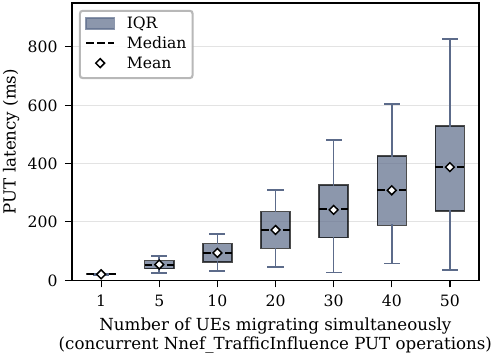}
  \caption{Per-subscriber \texttt{Nnef\_TrafficInfluence}
    PUT latency under concurrent load. $N$ simultaneous
    PUTs (one per SUPI) for $N \in \{1, 5, 10, 20, 30,
    40, 50\}$, ten repetitions each (1{,}560 operations,
    all HTTP~200). Boxes show median/IQR; whiskers
    1.5$\times$IQR.}
  \label{fig:concurrent_migrations}
\end{figure}
Fig.~\ref{fig:link_layer_latency} establishes that SubEdge
migration reduces \emph{link-layer} round-trip latency.
The complementary question---whether the AI inference
service the subscriber actually \emph{experiences}
remains continuous across the same migration events---is
the application-layer claim that distinguishes
SubEdge's integrated communication-and-compute service
from a pure traffic-steering mechanism. We replace the
ICMP probe with a 30\,Hz HTTP/POST inference workload
running through \texttt{uesimtun0} and re-execute the
same 600\,s mobility scenario. Each delivered frame is
logged with the edge node that served it (returned in
the response body), and frames are bucketed against
their scheduled wall-clock second so each 1\,s bucket
contains exactly 30 scheduled frames and the per-bucket
delivered count is bounded in $[0,30]$.

Fig.~\ref{fig:inference_continuity} mirrors the layout
of Fig.~\ref{fig:link_layer_latency} but reports the
application-layer metric in the top panel. The trace
holds at the 30\,fps target for 594 of 600 seconds. The
remaining six seconds correspond exactly to the six
migration events, with delivery dipping to 28\,fps in
three events and 27\,fps in the other three; in every
case the rate returns to 30\,fps in the immediately
following second. Across the run, \textbf{17{,}985 of
18{,}000 frames were delivered (99.92\%)}, and all 15
losses cluster within the 5GC policy-chain convergence
interval following each migration. No frame loss
occurred during stable operation on either edge node.
The mean trigger-to-first-frame convergence latency is
446\,ms, worst-case 607\,ms, dominated by the
single-process free5GC NEF's request handling. Together
with Fig.~\ref{fig:link_layer_latency}, this establishes
two-layer evidence that the same mechanism reduces link
RTT and delivers continuous AI inference to the moving
subscriber.

\subsubsection{Control-Plane: Concurrent Per-Subscriber
Migration}

Fig.~\ref{fig:concurrent_migrations} validates the control-plane
mechanism that drives per-subscriber migration---the
\texttt{Nnef\_TrafficInfluence} PUT that binds each
subscriber's inference traffic to the nearer edge
node---under concurrent mobility load. When $N$
AI-terminal subscribers move simultaneously, SubEdge
issues $N$ such PUTs concurrently, one per distinct
SUPI. All 1{,}560 PUT operations across $N \in \{1, 5,
10, 20, 30, 40, 50\}$ and ten repetitions returned
HTTP~200, confirming that the architecturally novel
step (the per-subscriber binding update via NEF) scales correctly under concurrent subscriber mobility.
Per-subscriber latency grows linearly with $N$ (slope
7.4\,ms, $R^2=0.999$), reflecting sequential request
processing in the single-process free5GC prototype NEF
rather than serialisation within SubEdge, which issues
all $N$ requests simultaneously. In a production
deployment the NEF is a cloud-native network
function (NF) scaled horizontally; per-subscriber
latency would remain at the single-subscriber
baseline of $\sim$20\,ms for any $N$, and the
architectural mechanism scales to thousands of
concurrently migrating subscribers per AF instance
through standard horizontal NEF replication.

\subsection{Limitations}

Two scope limitations bound the current
evaluation. First, free5GC v4.0.1 does not yet expose
\texttt{Nnef\_EventExposure}, so the testbed monitors
path latency directly; the migration mechanism this
triggers is identical to the AMF-driven path, and
substituting the trigger source is a configuration
change once the event becomes available. Second, the
testbed uses two edge nodes and one mobile UE,
sufficient to validate the per-subscriber mechanism
end-to-end but not to characterise placement
decisions across larger node populations; metropolitan-scale evaluation of the Placement Engine's selective re-evaluation policy is left for future work.

\section{Conclusion}
\label{sec:conclusion}

This paper has presented SubEdge, a subsystem that
enables mobile operators to provision integrated
communication and compute on a per-subscriber basis
to AI-native terminals. SubEdge contributes the
\emph{computing context} as a SUPI-to-container
binding primitive and a mobility-event-driven
mechanism that simultaneously migrates a per-subscriber compute instance and its per-subscriber
traffic-routing policy when the subscriber's serving
cell changes. Experimental evaluation on a real-world testbed running unmodified free5GC v4.0.1
shows that SubEdge's mobility-driven joint
communication-and-compute migration reduces 95th-percentile latency from 22.9\,ms to 12.2\,ms with
zero packet loss across six mobility events,
sustains 99.92\% frame delivery for an end-to-end
30\,fps inference workload, and completes 1{,}560 migration operations across batches of up to 50 simultaneously migrating subscribers with 100\%
success. SubEdge is to edge AI computing what IMS is
to multimedia sessions: a subsystem through which
operators extend their service model from
provisioning connectivity alone to provisioning both
connectivity and individualised inference compute, deployable on 5G infrastructure and positioned to evolve toward direct integration in 6G.
Future work targets two directions: scaling the
testbed to metropolitan topologies for evaluating
the Placement Engine at scale; and exploring
AI-based optimisation of per-subscriber joint
compute and communication migration, including
agentic AI policies that coordinate placement and
traffic-routing decisions across many concurrently
moving subscribers.

\bibliographystyle{IEEEtran}
\bibliography{references}

\end{document}